%% file: text.tex
\documentclass[twoside]{article}
\usepackage{fleqn,espcrc2}
\usepackage{graphicx}

\newcommand{\AmS}{{\protect\the\textfont2
  A\kern-.1667em\lower.5ex\hbox{M}\kern-.125emS}}
\newcommand{\D}{{\rm d}}
\newcommand{\mat}[3]{\left<{#1}\left|{#2}\right|{#3}\right>}
\newcommand{\pp}{p^\prime}
\newcommand{\RR}{\mathcal{R}}
\newcommand{\MM}{\mathcal{M}}
\newcommand{\xbar}{\overline{x}}
\newcommand{\meas}[2]{\frac{d^{#2}{#1}}{(2\pi)^{#2}}}

\title{Fracture functions in the very forward limit}

\author{B. E. White\address{Physics Department,
   University of Wales Swansea,\\
   Singleton Park,
   Swansea SA2 8PP, UK.}%
  \thanks{Talk presented at QCD99 Euroconference at Montpellier. Work
supported by PPARC. Preprint SWAT/99-237}}

\begin{document}

\pagestyle{empty}

\begin{abstract}
This talk gives a brief discussion of extended fracture functions,
which parametrise the non-perturbative physics in the target fragmentation
region of semi-inclusive DIS. In the forward limit $z \rightarrow 1$, it
can be seen that fracture functions can be identified with insertions of
composite operators. This enables polarised fracture functions to be
used to test a target-independence hypothesis of the ``proton spin effect''.
\end{abstract}

\maketitle

\section{Introduction}

The ``proton spin effect''~\cite{EMC} is the anomalous suppression of the 
Ellis-Jaffe sum-rule for the first moment of $g_1(x, Q^2)$ in inclusive
Deep Inelastic Scattering (DIS).
It can be understood in terms of a target-independent topological charge
screening mechanism~\cite{NSV}. Central to this analysis is the need to express
the moments of the target distribution functions as matrix elements of local,
composite operators which appear in the operator product expansion (OPE).

In a recent proposal~\cite{SV}, a suggestion was made to test the target-%
independence hypothesis in the target fragmentation region of semi-inclusive
DIS where a hadron is tagged in the final state.
This process factorises at large $Q^2$ as follows:
\begin{center}\begin{tabbing}
  Semi-incl\=usive structure function = \\
  \>(Hard physics) $\otimes$ (Fracture function).
\end{tabbing}\end{center}
Unlike parton densities in inclusive DIS, fracture functions cannot be related
to Green functions of local, composite operators. This prevents the analysis
of Refs.~\cite{NSV} to be applied rigorously to polarised, semi-inclusive DIS.
Instead, the moments of fracture function are represented as generalised,
space-like cut vertices~\cite{GTV}. 
However, in the forward
limit, $z \rightarrow 1$, where the tagged final state hadron
carries most of the nucleon momentum,
it can be shown~\cite{GSW} that these cut vertices reduce to objects depending
on insertions of local, composite operators.
This talk summarises the main arguments.

\section{Matrix elements and proper vertices}

This section summarises the explanation of the EMC/SMC ``proton spin''
effect as one of topological charge screening~\cite{NSV}. The important point
relevant to this analysis
is that the target distribution functions must be expressed in terms of
matrix elements of local, composite operators.

Measurements by the EMC and SMC collaborations~\cite{EMC} of the first moment~
$\Gamma^p_1$ of the polarised structure function~$g_1(x, Q^2)$ have found
it to be suppressed compared with its OZI expectation. The OPE for
$\Gamma^p_1$ is
\begin{eqnarray} \Gamma^p_1 & \;\; = \;\; & \int_0^1 dx\; g^p_1(x, Q^2)
   \nonumber \\ & =  &
   \frac{1}{12} C^{\rm NS}_1(\alpha_s) \left(a^3 + \frac{1}{3}a^8\right)
   \nonumber \\ & + & 
   \frac{1}{9}C^{\rm S}_1(\alpha_s) a^0(Q^2),
\end{eqnarray}
where the axial charges $a^i$ are defined as reduced matrix elements:
\begin{eqnarray}
   \frac{1}{2} \; a^3 \; s_\mu  & \; =\; & \mat{p;s}{A^3_\mu}{p;s} \nonumber \\
   \frac{1}{2\sqrt{3}} \; a^8 \; s_\mu & = & \mat{p;s}{A^8_\mu}{p;s} \nonumber
   \\
   a^0(Q^2) &  =  & \mat{p;s}{A^0_\mu}{p;s}
\end{eqnarray}
Note that the flavour singlet $a^0$ is scale-dependent because of the axial
anomaly. In the QCD parton model~\cite{BFR},
\begin{eqnarray}
  a^3 & = & \Delta u - \Delta d \nonumber \\
  a^8 & = & \Delta u + \Delta d - 2\Delta s \nonumber \\
  a^0(Q^2) & = & \Delta u + \Delta d + \Delta s - n_f \frac{\alpha_s}{2\pi}\Delta g(Q^2)
\end{eqnarray}
On the assumption of the OZI rule, $\Delta s \simeq 0 \simeq \Delta g$,
$a^0 \simeq a^8$, but experimentally 
$a^0$ is found to be suppressed compared with $a^8$.
This is not surprising given the scale-dependence of $a^0$; any explanation
of the suppression should take account of this.

One such explanation, given in a series of papers by Narison, Shore and
Veneziano~\cite{NSV}, is to identify the mechanism with topological charge
screening. Using the anomalous Ward identity
\begin{equation}
   \partial_\mu A^\mu_0 - 2n_f Q \simeq 0,
\end{equation}
with $Q = \frac{\alpha_s}{8\pi}\;{\rm tr}G_{\mu\nu}\widetilde{G}^{\mu\nu}$
the topological charge density, one can re-express the flavour singlet axial
charge as a measure of topological charge:
\begin{equation}
   a^0(Q^2) \;=\; \frac{1}{2M_p} 2n_f\mat{p}{Q}{p}
\end{equation}
We can then perform
a Legendre transform on the QCD generating functional with respect
to $Q$ and $\Phi_5 \propto \overline{q}\gamma_5 q$ only; this enables
a decomposition of $a^0$ into products of 1PI vertices and propagators.
\begin{eqnarray}
  a^0(Q^2) & = & \frac{1}{2M_p} 2n_f \left[
  \mat{0}{T(Q\,Q)}{0}\;\Gamma^{\rm 1PI}_{Q{\rm pp}} \right. \nonumber \\
  & & \phantom{2M_p 2n_f} \left. +
  \mat{0}{T(Q\,\Phi_5)}{0}\;\Gamma^{\rm 1PI}_{\Phi_5{\rm pp}}\right]
  \nonumber \\
  & = & \frac{1}{2M_p} 2n_f \left[
  \chi(0)\;\Gamma^{\rm 1PI}_{Q{\rm pp}}\right. \nonumber \\
  & & \phantom{2M_p 2n_f}
  \left. + \sqrt{\chi^\prime(0)}\;\Gamma^{\rm 1PI}_{\Phi_5{\rm pp}} \right],
\end{eqnarray}
where
\begin{eqnarray}
  \chi(k^2) & = & i\int d^dx\; e^{ik\cdot x} \mat{0}{T(Q(x)Q(0))}{0},
  \nonumber \\
  \chi^\prime(0) & = & \left. \frac{d}{dk^2}\chi(k^2) \right|_{k^2 = 0}.
\end{eqnarray}

From the chiral Ward identities, $\chi(0)$ vanishes in the chiral limit.
A similar decomposition for $a^8$ can be made in terms of $\Gamma_{\Phi^8_5}$.
Assuming~\cite{NSV} $\Gamma_{\Phi^8_5}$ obeys the OZI rule, it is found that

\begin{equation}
  \frac{a^0(Q^2)}{a^8} = \frac{\sqrt{6}}{f_\pi}\sqrt{\chi^\prime(0)}.
\end{equation}
Thus, the suppression of $a^0/a^8$ is due to an anomalously small value of
the slope of the topological susceptibility~$\chi^\prime (0)$. Furthermore,
this prediction does not depend on the nature of the target.

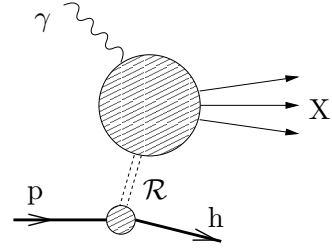
\begin{figure}\begin{center}
\input{diff_had.pstex_t}
\caption{Semi-inclusive DIS in the $z \rightarrow 1$ limit.}
\label{fig_diff_had}
\end{center}\end{figure}

To test this target-independence hypothesis, it has been proposed in Ref.~
\cite{SV} to measure the polarised structure functions of Regge poles
in semi-inclusive DIS in the limit $z \rightarrow 1$ where $z$ gives the
longitudinal momentum fraction of the final state hadron.
(Fig.~\ref{fig_diff_had}.) In this limit,
the polarised semi-inclusive structure function factorises:
\begin{eqnarray}
   \lefteqn{\sum_i e^2_i \Delta M_{\rm h/N}(x, z, t, Q^2)
   \stackrel{z \rightarrow 1}{\longrightarrow}} & & \nonumber \\
   & & F(t) (1-z)^{-2\alpha_\RR (t)}
   g^\RR_1(\frac{x}{1-z}, t, Q^2) 
\end{eqnarray}
By taking ratios of $\Delta M_{\rm h/N}$ for different final state hadrons
$h$ and targets $N = p,n$, we can effectively compare $g^\RR_1$
for different $\RR$. For example, the ratio of cross-sections
of processes with
$\RR = \Delta^\pm$ would be

\begin{equation}
   \frac{ {\rm e n} \rightarrow {\rm e}\pi^+{\rm X}}{
   {\rm e p} \rightarrow {\rm e}\pi^-{\rm X}} \simeq \frac{2s - 1}{2s + 2},
\end{equation}
where $s$ is a universal suppression factor measurable in inclusive DIS:

\begin{equation}
   s(Q^2) = \frac{C^S_1(\alpha_s)}{C^{NS}_1(\alpha_s)}\frac{a_0(Q^2)}{a^8}
\end{equation}

To put these predictions on a rigorous footing, we would need to show that one
can perform the OPE -- 1PI vertex -- propagator analysis sketched above
on semi-inclusive structure functions. The rest of this talk shows how this
is possible in the $z \rightarrow 1$ limit.

\section{Cut vertices and semi-inclusive DIS}

\begin{figure}\begin{center}
\input{gen_cut_vert.pstex_t}
\caption{Definition of generalised cut vertex.}
\label{fig_gen_cut_vert}
\end{center}\end{figure}
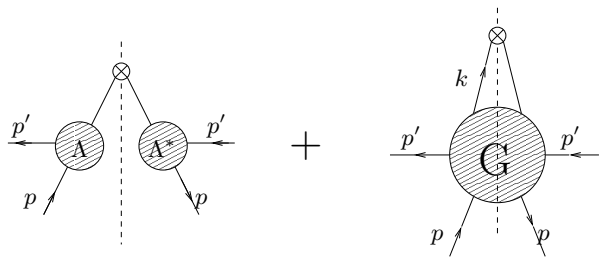

A simplified discussion of semi-inclusive DIS can be given in the language
of $(\phi^3)_6$ scalar field theory, which like QCD is asymptotically free.
To define a Lorentz-scalar semi-inclusive structure function, we can write
\begin{eqnarray}
  \lefteqn{W(q, p, \pp) =\frac{Q^2}{2\pi}\sum_X \int d^dx\;e^{iq\cdot x}}
  & & \nonumber \\
  & &  \phantom{****}\times \mat{p}{j(x)}{h,X} \mat{h,X}{j(0)}{p},
\end{eqnarray}
where $j(x) = \phi^2(x)$ plays the role of the electromagnetic current.
$q$ is the ``photon'' momentum, and $p$ and $\pp$ the proton and hadron $h$
momenta respectively. The convenient kinematical variables are

\begin{equation}\begin{array}{cc}
  Q^2 = -q^2, & t = (p - \pp)^2, \\
  z = \frac{\pp\cdot q}{p\cdot q}, & x = \frac{Q^2}{2p\cdot q}.
\end{array}\end{equation}
We also use $\xbar = x/(1-z)$. In the target fragmentation region, $Q^2$
large, $|t|/Q^2 \ll 1$, $W$ factorises into a convolution of perturbative
Wilson co-efficients $C$ and {\it extended} fracture functions
$\MM$~\cite{GTV}:
\begin{eqnarray}
  \lefteqn{W_{\rm targ} (Q^2, \xbar, z, t) =} & & \nonumber \\
  & & \int_{\xbar}^1 \frac{du}{u}
  \mathcal{M}(u, z, t, \mu) C(\frac{\xbar}{u}, Q^2, \mu),
\end{eqnarray}
Factorisation theorems have been proved for $(\phi^3)_6$ theory~\cite{GT}
and QCD~\cite{Col}. It has also been
shown~\cite{GT} that the moments of extended fracture functions~
$\MM^j$ can be represented as generalised, space-like, cut vertices:
(Fig.~\ref{fig_gen_cut_vert}.)
\begin{eqnarray}
   \mathcal{M}^j(z, t, \mu) & = & \int_0^1 \D u u^{j-1} \MM(u, z, t, \mu)
   \nonumber \\
   & = & (p_+ - \pp_+)^j |\Lambda(p, \pp)|^2 \nonumber \\
   & + &
   \int \meas{k}{d} (k_+)^j \theta(0 < k_+ < p_+) \nonumber\\
   & & \times\stackrel{\rm \textstyle disc}{\scriptstyle (k - p + \pp)^2}
   G(k,p,\pp),
\end{eqnarray}
where
\begin{eqnarray}
  \lefteqn{\Lambda(p,\pp) \; = \; \Delta_F^{-1}(p)\Delta_F^{-1}(\pp)} & &
  \nonumber \\
  & & \times \mat{0}{T(\phi(p)\phi(-\pp)\phi(-p + \pp))}{0}, \\
  \lefteqn{G(k,p,\pp) 
  \; = \;  \Delta_F^{-2}(p)\Delta_F^{-2}(\pp)} \nonumber \\
  & & \times \mat{0}{T(\phi_{p}\phi_{-\pp}\phi_{-k}\phi_{k}\phi_{\pp}\phi_{-p})}{0}.
\end{eqnarray}
(The first $\Lambda$-term is not required in QCD.)

The moments of fracture functions do not depend on the insertion of a composite
operator because of the final-state cut or discontinuity. Unlike in the
case of inclusive DIS~\cite{BFKM}, one cannot remove this final state cut ---
this is the reason why the OPE is not directly applicable to semi-inclusive
DIS. We cannot therefore
perform the 1PI vertex -- propagator decomposition on them.

\begin{figure}
  \centering{\input{dom_cuts.pstex_t}}
  \caption{Topology of the dominant one-loop diagrams. Compare
equation~\ref{dom_cuts}}
  \label{fig_dom_cuts}
\end{figure}
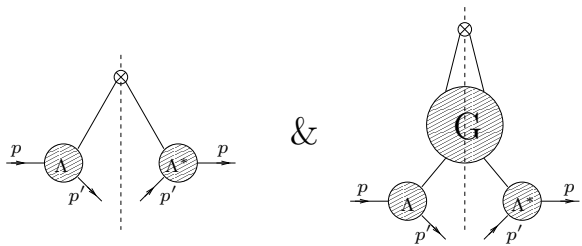

However, a one-loop
calculation in $(\phi^3)_6$ theory shows that the class of diagrams
contributing to $\MM^j$ which dominate as $z \rightarrow 1$ have such a simple
analytic structure that the final state cut can be removed. Their topology
is as in Fig.~\ref{fig_dom_cuts}.
(For details, see Ref.~\cite{GSW}.) The final result is
\begin{eqnarray}
   \lefteqn{\MM^j(z, t) \stackrel{z \rightarrow 1}{\longrightarrow}
   \left[ (p_+ - \pp_+)^j \;+\; \phantom{\meas{k}{d}} \right.} \nonumber \\
   & & \phantom{*}\left. \int\meas{k}{d}\;(k_+)^j G(k,p-\pp)\right]
   \Lambda^2(p,\pp) \nonumber \\
   & = & \mat{0}{T(\phi_{p-\pp}:\phi(i\partial_+)^j\phi:(0)\phi_{-p+\pp})}{0}
   \nonumber \\
   & & \times \Delta_F^{-2}(p-\pp) \Lambda^2(p,\pp)
\label{dom_cuts}\end{eqnarray}
where
\begin{eqnarray}
   \lefteqn{G(k_1, k_2) = \Delta_F^{-2}(k_1)}& & \nonumber \\
   & &
   \times\mat{0}{T(\phi(k_1)\phi(-k_2)\phi(k_2)\phi(-k_1))}{0}.
\end{eqnarray}
This leading term in $(1-z)$ depends on the Green function of a composite
operator.

The corresponding QCD cut vertices can also be expected to have this property.
As $z \rightarrow 1$, one can assume that factorisation of hard physics and
Regge factorisation~\cite{KS} are both valid, so that
\begin{eqnarray}
  \lefteqn{\MM^j_{a/h N}(z, t) \longrightarrow} \nonumber \\
  & & \sum_\RR F_{\RR N h}(t)
  (1 - z)^{-2\alpha_\RR(t)} f^j_{a/\RR}(t).
\end{eqnarray}
If this is the case, then the moments of Reggeon structure functions
$f^j_{a/\RR}$ do indeed depend on insertions of composite operators.

Having related fracture functions to insertions of composite operators, one
can then perform the 1PI vertex -- propagator decomposition on them. In the
spin-polarised case, this enables us to use fracture functions as a testing
ground for the target-independence hypothesis of the ``proton spin effect''.

\section*{Acknowledgements}

The work summarised in the last section~\cite{GSW} is co-authored with 
M. Grazzini and G. M. Shore. This author wishes to thank PPARC for a
research studentship, and also the organisers of the conference.

\end{document}

%% file: diff_had.pstex_t
\begin{picture}(0,0)%
\includegraphics{diff_had.pstex}%
\end{picture}%
\setlength{\unitlength}{0.00050000in}%
\begingroup\makeatletter\ifx\SetFigFont\undefined%
\gdef\SetFigFont#1#2#3#4#5{%
  \reset@font\fontsize{#1}{#2pt}%
  \fontfamily{#3}\fontseries{#4}\fontshape{#5}%
  \selectfont}%
\fi\endgroup%
\begin{picture}(3108,2539)(2368,-5044)
\put(2551,-4561){\makebox(0,0)[lb]{\smash{\SetFigFont{11}{13.2}{\familydefault}{\mddefault}{\updefault}p}}}
\put(5476,-3736){\makebox(0,0)[lb]{\smash{\SetFigFont{11}{13.2}{\familydefault}{\mddefault}{\updefault}X}}}
\put(3751,-4561){\makebox(0,0)[lb]{\smash{\SetFigFont{11}{13.2}{\familydefault}{\mddefault}{\updefault}$\mathcal{R}$}}}
\put(4426,-4861){\makebox(0,0)[lb]{\smash{\SetFigFont{11}{13.2}{\familydefault}{\mddefault}{\updefault}h}}}
\put(2626,-2761){\makebox(0,0)[lb]{\smash{\SetFigFont{11}{13.2}{\familydefault}{\mddefault}{\updefault}$\gamma$}}}
\end{picture}

%% file: gen_cut_vert.pstex_t
\begin{picture}(0,0)%
\includegraphics{gen_cut_vert.pstex}%
\end{picture}%
\setlength{\unitlength}{0.00041700in}%
\begingroup\makeatletter\ifx\SetFigFont\undefined%
\gdef\SetFigFont#1#2#3#4#5{%
  \reset@font\fontsize{#1}{#2pt}%
  \fontfamily{#3}\fontseries{#4}\fontshape{#5}%
  \selectfont}%
\fi\endgroup%
\begin{picture}(7449,3174)(1189,-4948)
\put(2026,-3661){\makebox(0,0)[lb]{\smash{\SetFigFont{9}{10.8}{\familydefault}{\mddefault}{\updefault}$\Lambda$}}}
\put(4801,-3661){\makebox(0,0)[lb]{\smash{\SetFigFont{17}{20.4}{\familydefault}{\mddefault}{\updefault}+}}}
\put(8251,-3511){\makebox(0,0)[lb]{\smash{\SetFigFont{9}{10.8}{\familydefault}{\mddefault}{\updefault}$\pp$}}}
\put(7951,-4711){\makebox(0,0)[lb]{\smash{\SetFigFont{9}{10.8}{\familydefault}{\mddefault}{\updefault}$p$}}}
\put(6226,-3511){\makebox(0,0)[lb]{\smash{\SetFigFont{9}{10.8}{\familydefault}{\mddefault}{\updefault}$\pp$}}}
\put(6601,-4711){\makebox(0,0)[lb]{\smash{\SetFigFont{9}{10.8}{\familydefault}{\mddefault}{\updefault}$p$}}}
\put(6901,-2761){\makebox(0,0)[lb]{\smash{\SetFigFont{9}{10.8}{\familydefault}{\mddefault}{\updefault}$k$}}}
\put(1276,-3361){\makebox(0,0)[lb]{\smash{\SetFigFont{9}{10.8}{\familydefault}{\mddefault}{\updefault}$\pp$}}}
\put(3751,-3361){\makebox(0,0)[lb]{\smash{\SetFigFont{9}{10.8}{\familydefault}{\mddefault}{\updefault}$\pp$}}}
\put(1426,-4261){\makebox(0,0)[lb]{\smash{\SetFigFont{9}{10.8}{\familydefault}{\mddefault}{\updefault}$p$}}}
\put(3601,-4261){\makebox(0,0)[lb]{\smash{\SetFigFont{9}{10.8}{\familydefault}{\mddefault}{\updefault}$p$}}}
\put(7201,-3886){\makebox(0,0)[lb]{\smash{\SetFigFont{17}{20.4}{\familydefault}{\mddefault}{\updefault}G}}}
\put(3001,-3661){\makebox(0,0)[lb]{\smash{\SetFigFont{9}{10.8}{\familydefault}{\mddefault}{\updefault}$\Lambda^*$}}}
\end{picture}

%% file: dom_cuts.pstex_t
\begin{picture}(0,0)%
\includegraphics{dom_cuts.pstex}%
\end{picture}%
\setlength{\unitlength}{0.00033300in}%
\begingroup\makeatletter\ifx\SetFigFont\undefined%
\gdef\SetFigFont#1#2#3#4#5{%
  \reset@font\fontsize{#1}{#2pt}%
  \fontfamily{#3}\fontseries{#4}\fontshape{#5}%
  \selectfont}%
\fi\endgroup%
\begin{picture}(9024,3741)(589,-3490)
\put(7576,-1861){\makebox(0,0)[lb]{\smash{\SetFigFont{14}{16.8}{\familydefault}{\mddefault}{\updefault}G}}}
\put(6076,-2686){\makebox(0,0)[lb]{\smash{\SetFigFont{7}{8.4}{\familydefault}{\mddefault}{\updefault}$p$}}}
\put(6976,-3436){\makebox(0,0)[lb]{\smash{\SetFigFont{7}{8.4}{\familydefault}{\mddefault}{\updefault}$\pp$}}}
\put(9301,-2686){\makebox(0,0)[lb]{\smash{\SetFigFont{7}{8.4}{\familydefault}{\mddefault}{\updefault}$p$}}}
\put(8401,-3436){\makebox(0,0)[lb]{\smash{\SetFigFont{7}{8.4}{\familydefault}{\mddefault}{\updefault}$\pp$}}}
\put(8476,-2986){\makebox(0,0)[lb]{\smash{\SetFigFont{7}{8.4}{\familydefault}{\mddefault}{\updefault}$\Lambda^*$}}}
\put(676,-2086){\makebox(0,0)[lb]{\smash{\SetFigFont{7}{8.4}{\familydefault}{\mddefault}{\updefault}$p$}}}
\put(1576,-2836){\makebox(0,0)[lb]{\smash{\SetFigFont{7}{8.4}{\familydefault}{\mddefault}{\updefault}$\pp$}}}
\put(3901,-2086){\makebox(0,0)[lb]{\smash{\SetFigFont{7}{8.4}{\familydefault}{\mddefault}{\updefault}$p$}}}
\put(3001,-2836){\makebox(0,0)[lb]{\smash{\SetFigFont{7}{8.4}{\familydefault}{\mddefault}{\updefault}$\pp$}}}
\put(5026,-1936){\makebox(0,0)[lb]{\smash{\SetFigFont{14}{16.8}{\familydefault}{\mddefault}{\updefault}\&}}}
\put(1351,-2386){\makebox(0,0)[lb]{\smash{\SetFigFont{7}{8.4}{\familydefault}{\mddefault}{\updefault}$\Lambda$}}}
\put(3076,-2386){\makebox(0,0)[lb]{\smash{\SetFigFont{7}{8.4}{\familydefault}{\mddefault}{\updefault}$\Lambda^*$}}}
\put(6751,-2986){\makebox(0,0)[lb]{\smash{\SetFigFont{7}{8.4}{\familydefault}{\mddefault}{\updefault}$\Lambda$}}}
\end{picture}